\documentclass[11pt]{article}
\usepackage{fullpage}
\usepackage{enumitem}
\usepackage{fullpage,xspace,amsmath,amsfonts,amssymb}
\usepackage{times,mathptmx}
\usepackage{graphicx,color}
\usepackage{amsfonts,amsmath,color,float,graphicx,verbatim, times}
\usepackage{amsthm}
\usepackage{multicol}
\usepackage{wrapfig}
\usepackage[margin=0.7in]{geometry}

\newtheorem{theorem}{Theorem}
\theoremstyle{plain}

\newtheorem{definition}{Definition}

\numberwithin{equation}{section}
\newif\ifiscolor
\iscolorfalse
\begin{document}
\long\def\/*#1*/{}
\pagestyle{empty}
\title{Hardness of Peeling with Stashes}
\author{Michael Mitzenmacher\thanks{Harvard University, School of Engineering and Applied Sciences. Supported in part by NSF grants CCF-0915922, IIS-0964473, and CNS-1011840.} \and Vikram Nathan}
\date{}
\maketitle
\thispagestyle{empty}

\begin{abstract}
The analysis of several algorithms and data structures can be framed
as a {\em peeling process} on a random graph: vertices with
degree less than $k$ and their adjacent edges are removed until no vertices of degree less than $k$ are left.  Often the question is
whether the remaining graph, the $k$-core, is empty
or not.  In some settings, it may be possible to remove either
vertices or edges from the graph before peeling, at some cost.
For example, in hashing applications where keys correspond to edges
and buckets to vertices, one might use an additional side data
structure, commonly referred to as a stash, to separately handle some
keys in order to avoid collisions.  The natural question in such cases
is to find the minimum number of edges (or vertices) that need to be
stashed in order to realize an empty $k$-core.

We show that both these problems are NP-complete for all $k \geq 2$, with the sole exception being that the edge variant of stashing is solvable in polynomial time for $k = 2$ on standard (2-uniform) graphs. 
\end{abstract}

\newpage
\twocolumn
\setcounter{page}{1}
\pagestyle{plain}

\section{Introduction}
The following {\em peeling process} can be used to find the $k$-core of a hypergraph: vertices with degree less than $k$ are
repeatedly removed, together with their associated edges. The $k$-core
is easily shown to be the maximal subgraph where each vertex has
degree at least $k$; it is therefore uniquely defined and does not
depend on the order vertices are removed in the peeling process.
Peeling processes, and variations on it, have found applications in
low-density parity-check codes \cite{LMSS,MV}, hash-based sketches
\cite{Bloomier,EG,GM}, satisfiability of random boolean formulae
\cite{BFU,mitzenmacher1997tight,Molloy}, and cuckoo hashing \cite{DGMMPR,PR}.  Usually in the design
of these algorithms the primary question is whether or not the $k$-core is
empty, and an empty $k$-core corresponds to a success.  We say that a hypergraph
is $k$-{\em peelable} if it has an empty $k$-core.

If the $k$-core is not empty, a natural question to ask is how many
edges or vertices need to be removed to yield an empty $k$-core.  This
question may have algorithmic implications.  For example, consider a
multiple-choice hash table of the following form.  There are $n$ keys
and $m$ buckets; each key has $d$ possible buckets where it can be
placed; and each bucket can hold at most $k-1$ keys.  By associating
buckets with vertices and keys with hyperedges (each key being an edge
of its $d$ possible buckets), we see that peeling can naturally
provide an assignment of keys to buckets satisfying the constraints.
When a vertex is removed, the bucket obtains all the keys
corresponding to adjacent edges; if the peeling yields an empty
hypergraph, all keys have been placed.  In this setting, removing an
edge from the hypergraph before peeling can correspond to placing a
key into a separate structure, often referred to as a stash.  If a
suitably sized stash can be implemented, peeling can efficiently find
an assignment, leading to the question of how many edges need to be
removed so the remaining hypergraph is $k$-peelable.\footnote{We note
that peeling does not completely solve the problem of assigning keys
to buckets; for example, work on cuckoo hashing shows that one can
do substantially better after peeling on random graphs to match
additional keys to buckets.  However, peeling provides a quick way
to create an assignment, and for cuckoo hashing, peeling first maintains
optimality in terms of the number of keys assigned.  Because of
this, understanding the limits of peeling in this context appears worthwhile.}  Even
without such algorithmic implications, the minimum number of vertices or edges
to remove to obtain a $k$-peelable graph appears a natural and interesting graph
theoretic question.  

Here we prove that determining the minimum number of vertices, or the
minimum number of edges, that need to be stashed (removed from the
graph) so that a $d$-regular hypergraph is $k$-peelable is
NP-complete for any $k,d \geq 2$. The only exception is that, in the
case of edges, determining the minimal stash is solvable in polynomial
time for $k = d = 2$; that is, for 2-cores on standard graphs.  Although
we believe the case of $k = d = 2$ is well understood, we briefly review it
Section~\ref{polytime}.  Given the connection to stashes for cuckoo
hash tables, we call this class of problems {\em stash problems}.

We note that a similar problem was recently considered in \cite{BKLRS-anchoring}.  In
their variation, they look at the {\em anchoring
  problem}: given a budget $b$, find the subset $B$ of $b$ vertices such
that peeling the graph of vertices from $V-B$ of degree less than $k$
yields the maximum number of remaining edges.  That is, the chosen
subset $B$ cannot be peeled, and serves as an anchor for the $k$-core.
They show that when $k=2$, the problem is solvable in linear time, and
when $k \geq 3$ the problem is NP-hard and further is NP-hard to
approximate within an $O(n^{1-\epsilon})$ factor for any constant
$\epsilon$.  One way of viewing the anchoring problem is that it {\em
  adds} to the graph; for example, by replacing a vertex in $B$ by a
clique of size at least $k$ (and appropriately connecting edges), one
can guarantee that vertex is anchored.  In contrast, our goal 
in stash probelms is to {\em remove} vertices or edges from the graph.

In what follows, we define terms and briefly consider
the case $k= d = 2$.  
We then show the problem of finding the optimal stash size when stashing vertices
is NP-complete for $k \geq 2$  by a reduction from Vertex Cover.  We then
reduce the problem of finding the optimal stash size for vertices to the
problem for edges.

\section{Notation and Definitions}

Recall that we say that a hypergraph is $k$-peelable if it has an
empty $k$-core.  A \emph{k-vertex-stash} of a hypergraph is a subset
of vertices $V \subset G$ such that $G - V$ is $k$-peelable.  (Of
course removing a vertex also removes all adjancent edges.)
Similarly, a \emph{k-edge-stash} of a hypergraph is a subset of edges
$E \subset G$ such that $G - E$ is $k$-peelable. For fixed values $k$ and $d$, the decision problem
$k$-VERTEX-STASH$(G_d, s)$ (respectively $k$-EDGE-STASH$(G_d, s)$)
asks whether the $d$-regular hypergraph $G_d$ has a minimal
$k$-vertex-stash (respectively $k$-edge-stash) of size at most $s$.
We use $k$-VERTEX-STASH and $k$-EDGE-STASH where the meaning is clear.  
We note that we could also consider non-regular hypergraphs in this
framework as well, but since stashing on these graphs can easily be reduced to stashing on regular hypergraphs, we consider only regular
hypergraphs here.  We refer to 2-regular hypergraphs as standard
graphs for clarity and convenience.
When we say PROBLEM1 $\leq_P$ PROBLEM2, we mean that there is a
polynomial time reduction from PROBLEM1 to PROBLEM2. In most proofs, we do not explicitly say
the reductions can be done in polynomial time since their implementations are easily seen to be linear.

\section{2-EDGE-STASH}
\label{polytime}

For standard graphs, the 2-core is empty precisely when the
graph has no cycles.  It follows readily that for 2-EDGE-STASH the
minimum number of edges that need to be removed equals the minimum
number of edges that need to be removed so that the graph has no
cycles; this well-known quantity is the {\em cyclomatic number} of the
graph, $h(G) = |E| - |V| + (\text{\# connected components in } G)$.  We note $h(G)$ is easily computed in
polynomial time by starting with an empty graph, inserting the
edges of $G$ one at a time in any order, and setting aside any edge
that forms a cycle, incrementing $h(G)$ accordingly.  (Standard union-find algorithms can be used to test
for cycles.)  

As we show later, the corresponding vertex stash problem for standard graphs
is actually NP-complete.

\section{$k$-VERTEX-STASH is NP-Complete}

To start, it is clear that $k$-VERTEX-STASH is in NP for any
$d$-regular hypergraph as the certificate is just the $k$-vertex-stash.  Also, the
standard NP-complete Vertex Cover problem is the degenerate
problem 1-VERTEX-STASH.  That is, since a graph has an empty 1-core if
and only if it consists of a collection of isolated vertices, the smallest
number of vertices to remove for 1-VERTEX-STASH is the size of the minimum
vertex cover.  This connection suggests a reduction from VERTEX COVER.

\begin{definition} A vertex cover of $G$ is a set of vertices $V$ such that all edges in $G$ are adjacent to at least one vertex in $V$. The NP-complete decision problem VERTEX-COVER$(G, s)$ asks if $G$ has a vertex cover of size at most $s$.
\end{definition}

\begin{theorem} VERTEX-COVER $\leq_P$ $k$-VERTEX-STASH for $k \geq 2$. 
\begin{proof}
Given a hypergraph $G$ we construct a hypergraph $G'$ by adding vertices and edges to $G$ such that 
$V^*$ is a minimal vertex cover of $G$ if and only if $V^*$ is a minimal $k$-vertex-stash of $G'$. To create $G'$, we
replace every edge $(u,v)$ by a subgraph $C_k(u,v)$ with the following properties.
\begin{enumerate}
\item $u,v \in C_k(u,v)$.
\item Each vertex $w \in C_k(u,v)$ has degree at least $k$. 
\item If either $u$ or $v$ is removed, $C_k(u,v)$ has an empty $k$-core.
\item $C_k(u,v)$ has $k$ edges adjacent to $u$ and $k$ edges adjacent to $v$.
\item $C_k(u,v)$ has no edges connected to any vertex in $G$ apart from $u$ and $v$. 
\end{enumerate}
To show that such hypergraphs exist, we first provide the construction for standard graphs ($d = 2$). Given $u, v$,
we construct $C_k(u,v)$ by adding $k$ new vertices (that we number
from 1 to $k$ for convenience), with each of these vertices connected to both $u$ and $v$. 
Also, vertices $1$ and $k$ have $k - 1$
additional edges to vertices $2, \ldots k - 1$ but not to each
other, while vertices $2, \ldots, k -1$ have edges to all $k$ vertices.

The third property is satisfied since if $u$
or $v$ is removed, vertices 1 and $k$ will then have degree $k - 1$
and can be peeled, which reduces all other vertices to degree $k-2$,
allowing them to be peeled, which completely removes $C_k(u,v)$.
The other properties are clearly satisfied.

For $d \geq 2$, simply add $d - 2$ extra ``dummy'' vertices
that are adjacent to every hyperedge in $C_k(u,v)$. Since $C_k(u,v)$
has at least $k$ edges, these $d-2$ vertices have degree at least $k$.
All other properties are easily seen to be satisifed.

\ifiscolor
\else
\begin{figure}[htbp]
\centering
\includegraphics[width=\linewidth]{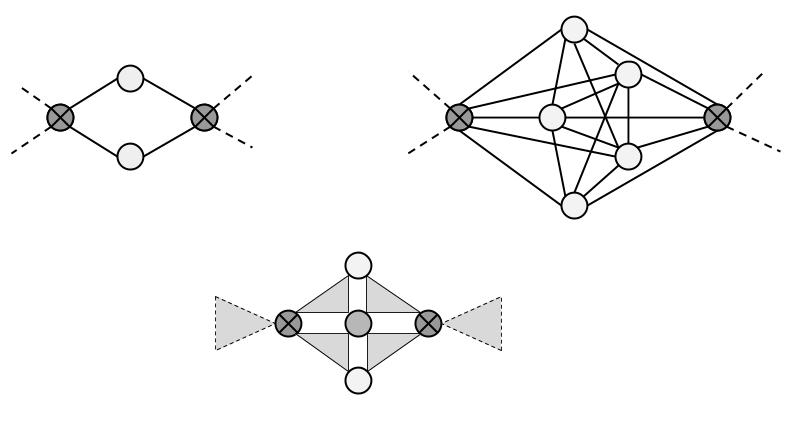}
\caption{\small Examples of $C_k(u,v)$ for various $d$ and $k$. $u$ and $v$ are gray and marked with an X. Top left: $k = 2, d = 2$. Top right: $k = 5, d = 2$. Bottom: $k = 2, d = 3$ with the dummy vertex in the center shaded darker. The hyper-edges are denoted by shaded triangles and join the three nodes they each touch.}
\label{ck-graphs}
\end{figure}
\fi

Now, we show that finding the minimal vertex cover of $G$ can be
reduced to finding the minimal vertex stash of $G'$.  To start, we
show that if $G'$ has a $k$-vertex-stash of size $s$, it has a
$k$-vertex-stash of size at most $s$ consisting entirely of vertices
from $G$; that is, we can repeatedly replace any $w \in C_k(u,v)$ in
the stash with either $u$ or $v$ until the stash consists only of
vertices from $G$.  To see this, let $S$ be a $k$-vertex-stash of $G'$
and suppose there exists $w \in S$ such that $w \in C_k(u,v)$ for some
$u,v \in G$. Define $S_v = \{v\} \cup S \setminus \{w\}$ and similarly
for $S_u$.  We claim that $S_v$ and $S_u$ are $k$-vertex-stashes of
$G'$.  Suppose that $S_v$ were not a $k$-vertex-stash. Then $G' -
S_v$ would have a non-empty $k$-core $T$ and since $S_v \cup \{w\}$ is a
$k$-vertex-stash, $w \in T$.  However, removing $v$ allows $C_k(u,v)$
to be peeled, giving a contradiction.  The same holds for $S_u$.
Hence if $G'$ has a $k$-vertex-stash of size $s$, it has a
$k$-vertex-stash of size at most $s$ consisting entirely of vertices
from $G$.

We can now show that $G$ has a vertex cover of size at most $s$ if and only if $G'$ has a $k$-vertex-stash of size at most $s$. 
In the forward direction, let $V^* \subset G$ be a vertex cover for $G$ with size
at most $s$. Since every edge in $G$ is adjacent to some $v \in V^*$,
$G' - V^*$ consists solely of the $C_k(u,v)$, each with either
or both of $u,v$ removed. Each of the $C_k(u,v)$
can then be peeled, so $G' - V^*$ has an empty $k$-core, and  hence $V^*$ is a
$k$-vertex-stash of $G$.

In the other direction, let $S \subset G'$ be a $k$-vertex-stash of
size at most $s$; as we have shown, without loss of generality, we can
assume $S \subset G$. Suppose that $S$ were not a vertex cover of
$G$. Then $G - S$ would contain edge $(u,v)$, so $G' - S$ would contain the
subgraph $\{u, v\} \cup C_k(u,v)$, which is not peelable, yielding a
contradiction. 

This completes the reduction.
\end{proof}
\end{theorem}

\section{$k$-EDGE-STASH is NP-complete}

We previously showed that 2-EDGE-STASH for standard graphs has a
polynomial time solution. We now show that $k$-EDGE-STASH is
NP-complete for both $k \geq 3$ (on all graphs) and $k = 2$, $d \geq
3$ by a reduction from $k$-VERTEX-STASH.  Of course it is clear that $k$-EDGE-STASH is in $NP$, since
one can check that a given stash leads to an empty $k$-core.  

In what follows, we use 
the following definitions: a \emph{neighbor} of a subgraph $S
\subset G$ is a vertex $v \notin S$ such that $v$ shares an edge with
some $w \in S$. A \emph{neighboring edge} of $S$ is any
edge shared by any neighbor of $S$ and some $w \in S$. We may often
refer to neighboring edges without specifying a precise neighbor; it
is understood that the neighbors will be specified later in a
construction that uses $S$. We call an edge \emph{internal} to $S$ if
it is in $S$.  Lastly, define a subgraph $S \subset G$ as $k$-\emph{unpeelable} if no vertex of $S$
has degree less than $k$, counting neighboring edges.

Throughout the reduction, vertices or edges may be \emph{peeled} or
\emph{stashed} (removed but not by the peeling process). These two
actions are mutually exclusive; however, we say an edge or vertex is
\emph{removed}, meaning either peeled or stashed, if the distinction
is irrelevant.

Finally, we have chosen to present the reduction by starting with a ``high-level'' argument, showing
the properties of the graph gadgets we require, and then subsequently showing how to construct actual gadgets with these properties (any construction satisfying these properties constitutes a valid gadget) We believe this approach is more enlightening than starting with the gadgets without motivation.

\begin{theorem} $k$-VERTEX-STASH $\leq_P$ $k$-EDGE-STASH for $k \geq 3$ and $k = 2, d\geq 3$.
\begin{proof}
Our main idea is to modify $G$ as follows:  given a hypergraph $G$, we construct a hypergraph $f(G)$ such that stashing a vertex $v \in G$ can be simulated by stashing an appropriate edge in $f(G)$. As a result, the size of the minimal vertex stash in $G$ will be equal to the size of the minimal edge stash in $f(G)$. We do this by replacing $v$ with a subgraph $P_k(v)$ so that this subgraph acts under peeling and edge stashing exactly as $v$ would act under peeling and vertex stashing, respectively.

To ensure that $P_k(v)$ correctly simulates $v$ in the peeling process, we need to be able to both (i) completely $k$-peel $P_k(v)$ in $f(G)$ exactly when $v$ can be $k$-peeled in $G$ and (ii) ensure that stashing and thus removing $v$ from $G$ can be simulated by stashing an edge in $P_k(v)$. This requires that $P_k(v)$ has some edge that can be stashed that allows the subgraph to be fully $k$-peeled. We formalize these properties as follows:
\begin{enumerate}[label=(P\arabic*)]
\item The degree of $v$ in $G$ is the number of neighboring edges of $P_k(v)$ in $f(G)$. Also, if $u$ is a neighbor of $v$, then $P_k(v)$ shares a neighboring edge with some vertex in $P_k(u)$. 
\item Provided that no internal edges are stashed, $P_k(v)$ can be $k$-peeled if and only if it has fewer than $k$ neighboring edges remaining (potentially after other removals of edges).  
\item There exists a non-empty set of internal edges \\ $E^*(P_k(v)) \subset P_k(v)$ such that $P_k(v)$ can be fully $k$-peeled if any $e \in E^*(P_k(v))$ is stashed.
\end{enumerate}
Notice that $f(G - \{v\}) = f(G) - P_k(v)$, and more generally: $f(G - V') = f(G) - \cup_{v \in V'} P_k(v)$. By (P3), $P_k(v)$ can be peeled by stashing a single edge, so this construction establishes an equivalence between stashing an edge in $f(G)$ and a vertex in $G$. Notice furthermore that $G$ can be $k$-peeled if and only if $f(G)$ can be $k$-peeled, which follows from observing that if there is no $v \in G$ with degree less than $k$, none of the $P_k(v)$ subgraphs can be peeled.

Let us assume for now that we can find such a $P_k(v)$. Then
we claim $G$ has a $k$-vertex stash $S$ of size at most $s$ if and only if $f(G)$ has a $k$-edge stash $S'$ of size at most $s$.
To prove the first direction, let $G_1 = G - S$, choose one $e_v^* \in E^*(P_k(v))$ for each $v$, and let $S' = \{e_v^* | v \in S \}$ be the corresponding edge stash. By (P3) of $P_k(v)$ above, stashing $S'$ in $f(G)$ allows all $P_k(v), v \in S$ to be $k$-peeled, resulting in $f(G) - \cup_{v \in S} P_k(v) = f(G_1)$. Since $G_1$ is $k$-peelable, $f(G_1)$ is $k$-peelable (with an edge stash of size $|S'| = |S|$).

To prove the other direction, let $S'$ be an edge stash of $f(G)$. Let $P' = \{v | \exists e \in S', e \in P_k(v)\}$ and note that $|P'| \leq |S'|$. Since $S' \subset \cup_{v \in P'} P_k(v)$, $f(G) - \cup_{v\in P'} P_k(v) = f(G - P')$ is $k$-peelable, so $G - P'$ is $k$-peelable with vertex stash $P'$. 

This completes the reduction assuming that $P_k(v)$ exists. In the following subsections, we show that a suitable $P_k(v)$ exists for both the cases $k \geq 3$, $d \geq 2$ and $k = 2, d \geq 3$.
\end{proof}
\label{main-edge-stash}
\end{theorem}

\subsection{Definitions of key components}

We define the key building blocks used in constructing $P_k(v)$ for both cases. 
\begin{definition}
A $b-$block is a $k-$unpeelable subgraph with $b$ neighboring edges that is $k-$peelable if any of its neighboring edges is removed.
\end{definition}

\begin{definition}
\label{def-stable-block}
A \emph{stable} block of degree $m$ is a $k-$unpeelable subgraph $B$ with the following properties: 
\begin{enumerate}
\item $B$ has $m$ neighboring edges. If no internal edge of $B$ is stashed, $B$ is $k-$peelable if and only if all $m$ neighboring edges are removed.
\item There exists a nonempty set $E^*(B) \subset B$ of internal edges such that $B$ is $k-$peelable upon stashing any $e \in E^*(B)$.
\end{enumerate}
\end{definition}

\subsection{Construction for $k \geq 3$ and $d = 2$}

To begin we assume that we have stable blocks and 3-blocks available, and then provide their construction.  
For vertex $v$, let $\delta = \deg(v)$. As shown in Figure~\ref{Pkv}, we can construct a $P_k(v)$ that consists of:
\begin{itemize}
\item A copy of vertex $v$, called the primary node.
\item $\delta$ instances of 3-blocks ($b$-blocks with 3 neighboring edges).
\item a stable block $B$, with $\delta$ neighboring edges.
\end{itemize}
$P_k(v)$ is constructed as follows: Let $\{u_i\}_{i = 1}^{\delta}$ be the neighbors of $v$, and index the 3-blocks in $P_k(v)$ as $b_{u_1}, \ldots b_{u_\delta}$. For each edge $(u_i, v)$, add 3 edges connecting $b_{u_i}$ to the primary node, stable block, and $b_v \in P_k(u_i)$.

\ifiscolor
\else
\begin{figure}[htbp]
\centering
\includegraphics[width=\linewidth]{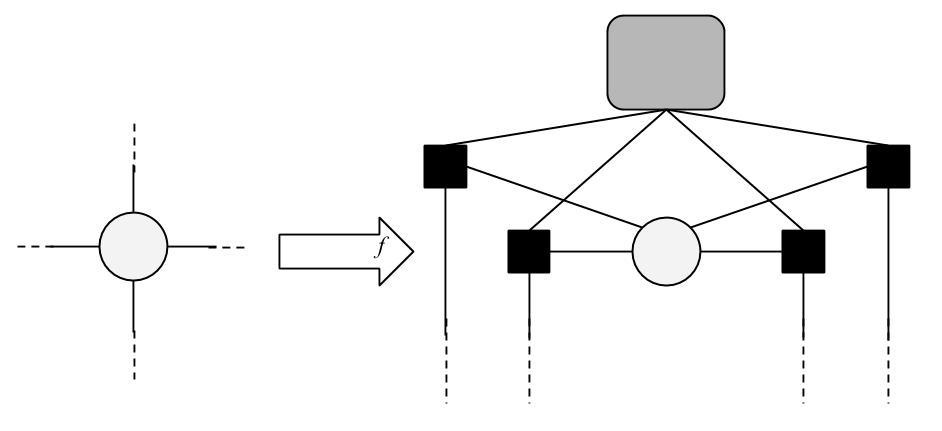}
\caption{\small The mapping (under $f$) of a single $v \in G$ to $P_k(v) \subset G'$. The black squares are 3-blocks, and the large gray structure is the stable block. Note that $k$ doesn't play an explicit role in this structure - it is hidden in the structure of the $b$-blocks and stable block.}
\label{Pkv}
\end{figure}
\fi

This construction of $P_k(v)$ satisfies the 3 required properties listed
above. (P1) is satisfied by construction. If $(u,v) \in G$, the edge between $P_k(v)$ and $P_k(u)$ is the edge between $b_u \in P_k(v)$ and $b_v \in P_k(v)$ as described above.

(P2) is ensured by the 3-blocks. Note that the primary node has
the same number of neighboring edges as $P_k(v)$ at all times, since each 3-block is connected to a neighboring
edge of $P_k(v)$ and is adjacent to the primary node. Thus, if any
neighboring edge is removed, the degree of the primary node will
decrease by 1 after peeling, and vice versa. We can use this to prove (P2): if $P_k(v)$ has fewer than $k$ neighboring edges, the primary node has degree less than $k$ and can be peeled, which peels all the $b_{u_i}$, leaving the stable block with no neighboring edges. Since stable blocks can be peeled completely if they have no neighboring edges, $P_k(v)$ can be fully peeled. Conversely, if $P_k(v)$ has at least $k$ neighboring edges, the primary node has degree at least $k$ and cannot be peeled. In this case, it is clear that $P_k(v)$ is unpeelable.

The stable block ensures (P3), that there is an internal edge
of $P_k(v)$ that can be stashed that allows $P_k(v)$ to be $k$-peeled.
Specifically, there is an edge that when stashed allows the stable
block to be $k$-peeled, and subsequently the $3$-blocks in turn are
$k$-peeeled.  Note that this means that $E^*(B) \subseteq
E^*(P_k(v))$.

We now detail the constructions of $b-$blocks and stable blocks.

\subsubsection{Constructing $b$-blocks}

\begin{figure}[htbp]
\centering
\includegraphics[width=\linewidth]{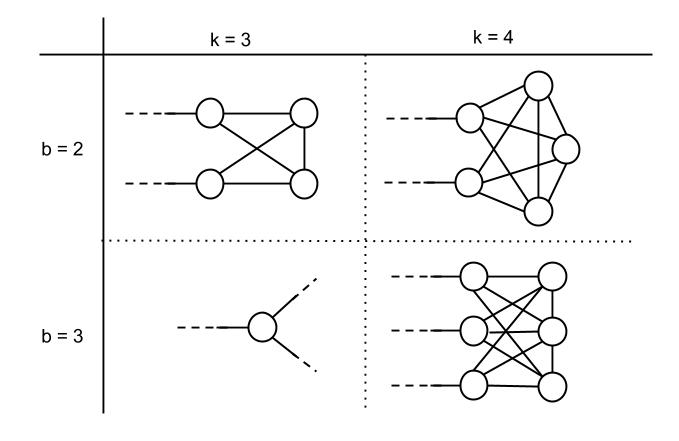}
\caption{\small Construction of $b-$blocks for $b = 2,3$ and $k = 3, 4$.}
\label{bblocks}
\end{figure}

While we note that $b$-blocks exist for every $b \geq 2$ and $k \geq 3$, we only require 2- and 3-blocks, so only we explicitly construct those.  An example is given in Figure~\ref{bblocks}. 
\begin{itemize}
\item A 2-block consists of two ``hub'' nodes, each adjacent to a neighboring edge, and a clique of size $k-1$. Both hub nodes have an edge to each node in the clique. Since every vertex in this block has degree $k$, it cannot be peeled. However, it's easy to see that if either neighboring edge is removed, the whole block can be peeled.
\item See Figure~\ref{bblocks} for 3-blocks with $k = 3,4$. Note the special case for $k = 3$, which is a single node.
For $k \geq 5$, a 3-block is created by first having 3 hub nodes, each adjacent to one of the neighboring edges, which we call ``Layer 0''. There are then $k-1$ nodes in ``Layer 1'' and a clique of size $k - 3$ as ``Layer 2''. Each hub node shares an edge with each Layer 1 vertex, and each Layer 1 vertex shares an edge with each Layer 2 vertex. If no edge is removed, the block cannot be peeled, since the vertices in Layers 0 and 1 have degree $k$, while the Layer 2 nodes have degree $2k-5 \geq k$ since $k\geq 5$. Notice that if any neighboring edge is removed, the corresponding hub node will have degree $k - 1$ and can be peeled, leaving all Layer 1 nodes with degree $k-1$, which after peeling leaves all degree 2 vertices with degree $k-4$, allowing the whole block to be peeled.
\end{itemize}

\subsubsection{Constructing simple stable blocks}

Before constructing stable blocks, we first construct what we call a
\emph{simple stable block}, which is a stable block with degree $m
\leq k-1$. It consists of a central vertex with degree $k - 1 + m$, two
2-blocks, and ($k-3$) 3-blocks. The central vertex connects to $m$
neighboring edges, and each of the remaining $k-1$ edges adjacent to
the central vertex are shared with one of the $k-1$ 2- or
3-blocks. The $b$-blocks themselves are connected: the $i$th $b$-block
shares edges with the $(i-1)$st and $(i+1)$st $b$-block, as shown in
Figure~\ref{k-1stable}. As a result, the 1st and $(k-1)$st $b$-blocks are
2-blocks, while the other $k-3$ of them are 3-blocks.

\ifiscolor
\else
\begin{figure}[htbp]
\centering
\includegraphics[width=\linewidth]{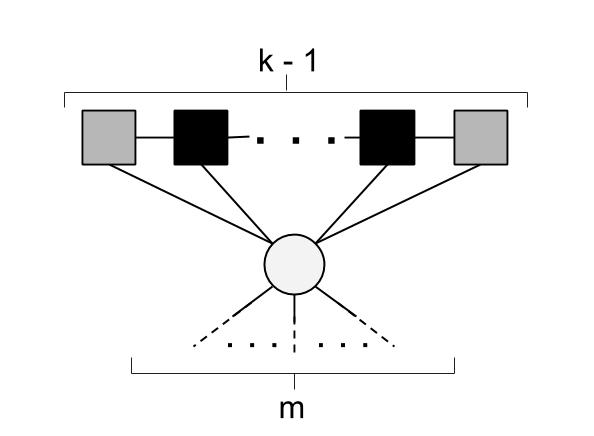}
\caption{\small Example of a simple stable block construction. Black squares are 3-blocks and gray squares are 2-blocks.}
\label{k-1stable}
\end{figure}
\fi

This satisfies the two properties of a stable block, since:
\begin{enumerate}
\item If at least one neighboring edge of the stable block remains and no internal edges are stashed, the central node has degree at least $k$ and cannot be peeled.
\item $E^*$ contains any edge adjacent to the 2- or 3-blocks. Removing any of these edges will remove adjacent $b$-blocks and then successively all the other $b$-blocks, since they are connected. This leaves the single central node, which would then have degree at most only $k-1$ by construction.
\end{enumerate}

\subsubsection{Constructing general stable blocks for $k \geq 3$}

Given the construction of stable simple blocks with degree $m \leq
k-1$, we can construct a stable block of arbitrary degree $m \geq k$
by using a $k-1$-ary tree of simple stable blocks, each of degree
$k-1$, constructed above. Each neighboring edge of the central vertex
of a simple stable block at depth $i$ is connected to a 2-block of a
simple stable block at depth $i+1$ (which requires converting said 2-block into a
3-block), so that the whole tree has at least $m$ neighboring
edges. Any excess neighboring edges at the base of the tree can be removed to give the tree
exactly $m$ neighboring edges. See
Figure~\ref{generalstable} for an illustration.

\ifiscolor
\else
\begin{figure}[htbp]
\centering
\includegraphics[width=\linewidth]{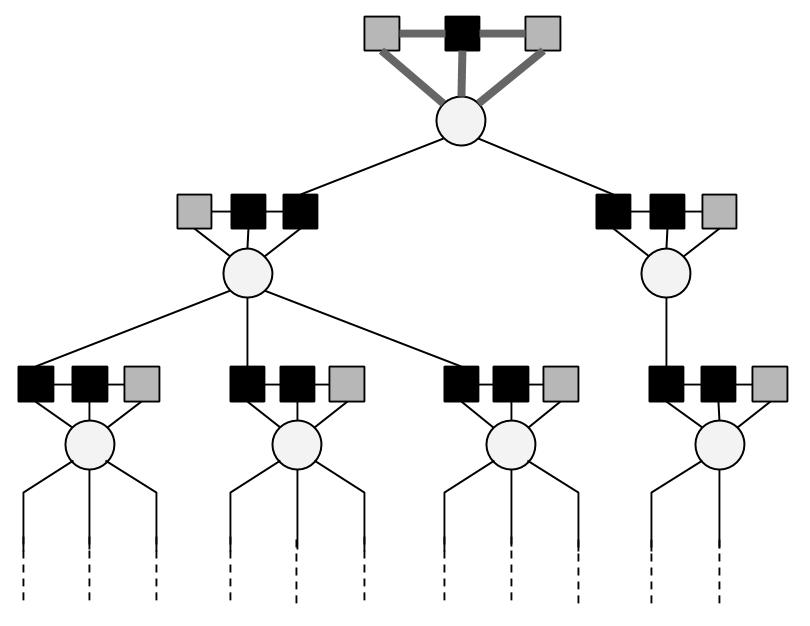}
\caption{\small Example of a general stable block, shown here with 11 neighboring edges for $k = 4$. Again, gray squares are 2-blocks, while black squares are 3-blocks. The thickened edges of the root node are all in $E^*$. Notice that the tree is made entirely out of simple stable blocks.}
\label{generalstable}
\end{figure}
\fi

This satisfies both the requirements of a stable block:
\begin{enumerate}
\item If a neighboring edge exists, there is at least one simple stable block at the base of the tree. 
It, and its parents, and so on up the tree, cannot be peeled assuming that no internal edges are stashed. 
It is also easy to see that the tree is then peeled if no neighboring edge remains.
\item If $R_s$ is the root simple stable block of this tree, then $E^*(R_s) \subseteq E^*($stable block).  Once the root simple stable block
is peeled, its children can be peeled, and so on for the whole stable block.
\end{enumerate}

This completes the construction of $P_k(v)$. If $v$ has degree $\delta$, then $P_k(v)$ has size $O(\delta k)$. 

\subsubsection{Adapting the construction for $d \geq 3$}
 
This construction can be easily adapted for $d \geq 3$ by simply
adding an additional $d - 2$ dummy vertices and extending each
internal edge in $P_k(v)$ to include these extra vertices. These
dummy vertices will not affect the peeling, since a dummy vertex
cannot be peeled until there are fewer than $k$ edges remaining in the original graph.

\subsection{Construction for $k = 2$ and $d \geq 3$}

For this case, we require a different construction; we do not use
$b$-blocks in our construction of the stable block, and we require a
slightly different construction for a suitable $P_2(v)$ building block
as well. At the heart of a stable block $T$ of degree $p$ is a
$d-1$-ary tree: a tree where each node has $d-1$ children all joined to it via
the same edge.  (Note this requires $d \geq 3$, and hence is consistent
with the fact that the case $d=k=2$ is polynomial time.)
This tree must have at least $p$ leaves. 
Suppose $v_1, \ldots, v_{d-1}$ are leaves of the tree connected
to the same parent by the same edge. Then extra nodes $w_1, \ldots,
w_{d-1}$ are created, in addition to the edges $(v_i, w_1, \ldots,
w_{d-1})$ for each $1\leq i\leq d-1$. See
Figure~\ref{stable-block-hypergraph}. Notice that every node of the
tree has degree at least 2, except the root $r$, which has degree
1. Notice also that if we were to remove the \emph{root edge} $e_r$
(the edge connecting the root to its immediate children in the tree),
the entire tree would be subsequently 2-peeled.

To complete the construction of the stable block $T$ of degree $p$, choose any $p$ of the $w_i$. For each $1 \leq i \leq p$, add a neighboring edge $(r, w_i, x_{i,1}, \ldots, x_{i,d-2})$, where the $x_{iq}\; (1 \leq q \leq d-2)$ are a collection of $p(d-2)$ new vertices and neighbors of the stable block.  We show below how the $x_{iq}$ are used to finish the construction of $P_k(v)$. Notice now that, with the neighboring edges, $\deg(r) = p+1$.

The above construction $T$ is a stable block for $k = 2$ and $d-$uniform hypergraphs, where $d \geq 3$. 
It satisfies the 2 properties of stable blocks:
\begin{enumerate}
\item Suppose that no internal edge of $T$ is stashed. Since every vertex in $T$, except $r$, has at least 2 internal edges, the only way for $T$ to be peeled is for $r$ to be peeled, which can happen only if all neighboring edges are removed. Thus, $T$ is $k$-unpeelable if it has degree at least 1.
\item If $e_r$ is stashed, then the entire $d-1$-ary tree can be subsequently peeled. Since each $w_i$ would then have degree only 1 (the neighboring edge), it can be peeled, peeling the neighboring edges as well. Thus $e_r \in E^*(T)$.
\end{enumerate}
As with the stable block in the previous construction for $k \geq 3$, the stable block is responsible for (P3) of $P_2(v)$, which allows stashing a single edge in $P_2(v)$ to correctly simulate stashing $v$ in $G$. 

Given this new stable block construction, construct $P_2(v)$ as follows. For every vertex $v \in G$, let $\delta = \deg(v)$ and let $\{e_i\}_{i = 1}^\delta$ be the edges in $G$ adjacent to $v$ and let $m_{ij}\;(1 \leq j \leq d-1)$ be the neighbors of $v$ in $G$ that share edge $e_i$ with $v$. Then $P_2(v)$ consists of:
\begin{itemize}
\item A copy of $v$ in $P_2(v)$, which is called the primary vertex and will serve the same role as the primary vertex in the previous $P_k(v)$ construction.
\item A stable block $T$ of degree $\delta$ (recall that this means $T$ has $\delta(d-2)$ neighbors). 
\item A vertex $t_{e_i}$ for each neighboring edge $e_i$.
\item Add $d-2$ dummy vertices for each $i$, and denote them by $x_{i1},\ldots,x_{i,d-2}$.
\end{itemize}
To combine these, add the edges $(v, t_{e_i}, x_{i1}, \ldots, x_{i,d-2})$ for each $i$.

 We further need to define the neighboring edges of $P_k(v)$. Note that each $P_k(m_{ij})$ will have its own copy of $t_{e_i}$ since all the $m_{ij}$ share edge $e_i$. Call this copy $t_{e_i}^{m_{ij}} \in P_k(m_{ij})$. Then, for each $t_{e_i} \in P_k(v)$, add the hyperedge $(t_{e_i}, t_{e_i}^{m_{i1}}, \ldots, t_{e_i}^{m_{i,d-1}})$. These constitute the neighboring edges of $P_2(v)$. See Figure~\ref{stable-block-hypergraph} for an illustration.

Each $t_{e_i}$ and $x_{iq}$ has degree 2, while the primary node has degree $\delta$. The $t_{e_i}$ in this construction are close analogs of the $b_{e_i}$ (the 3-blocks) used in the previous construction for $k \geq 3$, since both are responsible for ensuring (P2) of $P_k(v)$. Notice that the number of neighboring edges of $P_2(v)$ is always equal to the number of $t_{e_i}$s present, since if an edge of $P_2(v)$ is removed, the corresponding $t_{e_i}$ has degree 1 and can be peeled, which reduces the degree of the primary vertex $v$ by 1. As such, it ensures that the degree of the primary node is equal to the number of neighboring edges of $P_2(v)$, which means that the primary vertex can be peeled if and only if $P_2(v)$ has fewer than 2 neighboring edges. This is identical to the role of $b_{e_i}$ in the previous construction.

The dummy vertices are there simply to interface with the stable block. Since each edge must contain $d$ vertices, these dummy vertices just fill up the extra $d-2$ slots required for the stable block's neighboring edges to be complete.

\ifiscolor
\else
\begin{figure}
\centering
\includegraphics[width=\linewidth]{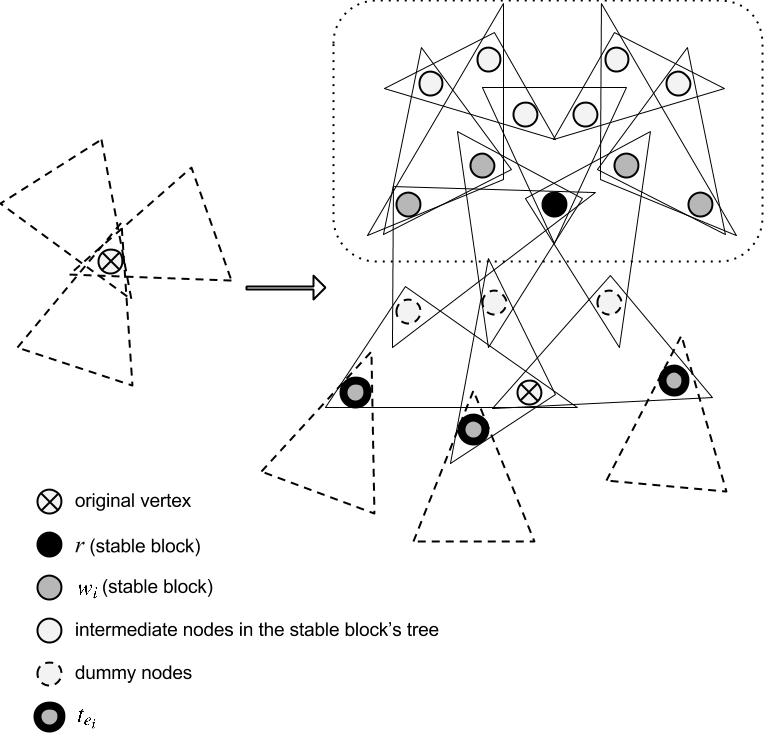}
\caption{\small An illustration how a vertex $v$ with degree 3 in $G$ is mapped to a set of vertices in $G'$ for $d = 3$. The stable block is the dotted box, with the root of the tree $r$ marked in black. The hyper-edges are shown as triangles containing all the nodes they connect.}
\label{stable-block-hypergraph}
\end{figure}
\fi

We now show that $P_2(v)$ satisfies the required properties.
(P1) is clearly satisfied by construction.
For (P2), if $P_2(v)$ has degree less than 2, then there are fewer than 2 $t_{e_i}$s, so $v$ can be peeled. Then each $t_{e_i}$ and $x_{ij}$ has degree at most 1 and can be peeled. The stable block then has degree 0 and is subsequently peeled, so $P_2(v)$ be peeled entirely.  Similarly, if $P_2(v)$ has degree at least $2$, then at least 2 $t_{e_i}$s exist, so $v$ has degree at least 2.  The stable block therefore has at least $2$ neighbors and cannot be peeled.
For (P3), the root edge $r$ of the $d-1$-ary tree is in $E^*(P_2(v))$. As shown above, stashing this edge peels the stable block, which peels the $x_{ij}$ and in turn the $t_{e_i}$s and $v$.

It is clear that $f(G)$ can be constructed in time polynomial in $|G|$.

\section{Conclusion}

We have shown that determining the minimum number of vertices (or
edges) that need to be removed from a hypergraph so that peeling leads
to an empty graph is NP-complete (except for $k=d=2$ for edges).  This
may potentially limit the use of stashes in algorithmic settings where
peeling could be a useful tool, although we would expect in many situations
non-optimal stashes based on reasonable heuristics would still prove effective in practice.

\end{document}